\documentclass[a4paper,twocolumn,
english,aps,pre,floatfix,showpacs]{revtex4}
\usepackage[T1]{fontenc}
\usepackage[latin1]{inputenc}
\usepackage{amsmath}
\usepackage{babel}
\usepackage{graphics}
\usepackage{amssymb}
\usepackage[dvips]{graphicx}
\usepackage{mathrsfs}
\usepackage{verbatim}

\begin{document}
\title
{Logarithmic corrections to correlation decay in two-dimensional random-bond Ising 
systems}
\author {Jean C. \surname{Lessa}}

\email{jean@if.ufrj.br}

\affiliation{Departamento de F\'{i}sica, Universidade Estadual de Feira de 
Santana, Campus Universit\'ario, 44031-460
Feira de Santana BA, Brazil}

\author {S.L.A. \surname{de Queiroz}}

\email{sldq@if.ufrj.br}

\affiliation{Instituto de F\'\i sica, Universidade Federal do
Rio de Janeiro, Caixa Postal 68528, 21941-972
Rio de Janeiro RJ, Brazil}

\date{\today}

\begin{abstract}
The statistics of critical spin-spin correlation functions in Ising
systems with non-frustrated disorder are investigated on a strip geometry,
via numerical transfer-matrix techniques. Conformal invariance concepts are
used, in order to test for logarithmic corrections to pure
power-law decay against distance. Fits of our data to conformal-invariance
expressions, specific to logarithmic corrections to correlations on strips,
give results with the correct sign, for the moments of order $n=0-4$ of
the correlation-function distribution.
We find an interval of disorder strength along which corrections to pure-system
behavior can be decomposed into the product of a known $n$-dependent factor and an 
approximately $n$-independent one, in accordance with predictions.
A phenomenological fitting procedure is proposed, which takes partial account of 
subdominant terms of correlation-function decay on strips. In the low-disorder 
limit, it gives results in fairly good agreement with theoretical predictions, provided
that an additional assumption is made. 
\end{abstract}
\pacs{PACS numbers: 05.50.+q, 75.10.-b}

\maketitle
 
\section{Introduction} 
\label{intro}
Magnetic systems with quenched disorder exhibit properties which often differ from 
those of their homogeneous counterparts. The Harris criterion~\cite{har74} for the 
relevance of disorder, with respect to possible changes of universality class at the
magnetic phase transition, boils down to the question of whether the pure-system
specific heat exponent $\alpha$ is positive or negative. In the former case, the
universality class changes, relative to that of the pure system, while in the latter it 
does not. For the two-dimensional Ising model, the Harris criterion is indecisive,
as the pure-system specific heat diverges logarithmically. Thus, alternative schemes
must be formulated in order to investigate this borderline case. Nowadays, mounting 
evidence~\cite{sha94,sst94} suggests that the critical 
behavior of two-dimensional Ising ferromagnets  is only slightly modified by 
the introduction of non-frustrated disorder. The corresponding changes are
given by logarithmic 
corrections to pure-system power-law singularities. Though the theoretical 
framework underlying such corrections is well-understood, their detection
in numerical studies turns out to be
a subtle matter~\cite{dQ97,sbl2,msbl,bs04}.

Upon introduction of disorder, the scaling dimensions associated to the moments of 
correlation functions at criticality acquire multifractal behavior~\cite{prb60_3428},
characterized by a set of independent exponents, as opposed to the pure case
where a single exponent is expected.

It was analytically predicted by Ludwig~\cite{npb330_639} that the moments of the 
spin-spin correlation functions of the disordered Ising model behave asymptotically as:
\begin{equation}
\left[\left\langle  \sigma_{0}\sigma_{R}\right\rangle _{J_{ij}}^{n}\right] \sim 
R^{-n/4}\left( \ln R\right) ^{n(n-1)/8}\ 
\label{eq:1}
\end{equation} 
where $n=1,2, \dots$, and the label $J_{ij}$ denotes average over quenched disorder 
(bond) configurations. The power-law term  $R^{-n/4}$ corresponds to the uniform-system
correlation decay, while the logarithmic corrections are given by 
$\left(\ln R\right) ^{n(n-1)/8}$. Critical correlation functions are 
non-self averaging quantities, that is, the width of their distribution stays 
essentially constant as the number of samples $N$ increases; however, the 
sample-to-sample fluctuations of averaged values (e.g., the assorted moments just 
mentioned) do go down approximately with the square root of sample size as the latter 
increases~\cite{dQrbs96}.

For the two-dimensional Ising model, the ``typical'', or most probable, correlation 
function, defined for a specific disorder configuration (i.e., the zeroth order moment 
of the distribution), is predicted~\cite{npb330_639} to behave as: 
\begin{equation}
G_0\equiv \exp\left[ \ln G\left( R\right) \right]_{\rm av}\sim
R^{-1/4}\left( \ln 
R\right) ^{-1/8},
\label{eq:2}
\end{equation}
where, again, the uniform-system behavior ($G(R)\sim R^{-\eta},\ \eta=1/4$) 
is accompanied by a logarithmic correction. Note that corrections are expected to be 
absent for $n=1$, i.e., for the average correlation decay.
For higher moments $n\geq 2$, the correction terms should be present,
with corresponding exponents $\lambda _{n}\equiv n(n-1)/8$, 
given in Eq.~(\ref{eq:1}).

Here we apply numerical transfer-matrix methods, together with finite-size scaling
and conformal invariance concepts, to investigate the random-bond Ising model on strips
of a square lattice ($L\times N,\ N\rightarrow \infty$). Our purpose is to
confirm the analytical results derived by Ludwig, for the exponents 
$\eta_{n}=n/4$ and $\lambda_{n}=n(n-1)/8$, associated to the moments of
critical correlation functions of the model.

\section{Method}
\label{method}
We consider a random-bond Ising system on a square lattice, whose Hamiltonian is given by
\begin{equation}
\mathscr{H}=-\sum_{i,j}J_{ij}\sigma_{i}\sigma_{j}\ ,
\label{eq3}
\end{equation} 
where $\sigma_{i}=\pm1$ are the (site) spin variables, and
$J_{ij}$ are ferromagnetic interactions between nearest-neighbor spins, extracted from 
the quenched probability distribution:
\begin{equation}
P(J_{ij})=\dfrac{1}{2}\left[ \delta(J_{ij}-J_{0})+\delta(J_{ij}-rJ_{0})\right], 
0\leqslant r \leqslant 1 .
\label{eq4}
\end{equation} 
The exact critical temperature $\beta_c\equiv1/k_{B}T_c$ is known from 
duality~\cite{jsp18_111, prb23_3421}
\begin{equation}
\sinh(2\beta_{c}J_{0})\sinh(2\beta_{c}rJ_{0})=1 .
\label{eq5}
\end{equation}
In our calculations we initially consider a system with  $r=1/4$, for which 
Eq.~(\ref{eq5}) gives $T_{c}(1/4)/J_{0}=1.239\cdots$ (in these units, the critical point 
of the uniform system is at $T_{c}(1)/J_{0}=2.269\cdots$). We shall also investigate 
stronger disorder, by using $r=0.1$, $0.05$, and $0.01$ (for the latter, the critical
temperature is $T_{c}(0.01)/J_{0}=0.5089\cdots$). 
The calculation of spin-spin correlation functions
follows the lines of Section 1.4 of Ref.~\onlinecite{fs2}, with standard adaptations 
for an inhomogeneous system\cite{dQ95}.
Taking two spins on, say, row 1, separated by a distance  $R$, and for a 
given configuration $\mathcal{C}$ of bonds, one has:
\begin{equation}
\langle \sigma_{0}^1 \sigma_{R}^1 \rangle_{\cal C} =
\frac{\sum_{\sigma_{0} \sigma_{R}}\tilde\psi 
(\sigma_{0})\, \sigma_{0}^1\ \left(\prod_{i=0}^{R-1} {\cal T}_{i}\right)_ 
{\sigma_{0} \sigma_{R}}\ \sigma_{R}^1\, \psi (\sigma_{R})} 
{\sum_{\sigma_{0} \sigma_{R}}  \tilde\psi (\sigma_{0})\ \left( 
\prod_{i=0}^{R-1} {\cal T}_{i}\right)_{\sigma_{0} \sigma_{R}}\ \psi 
(\sigma_{R})}\ \ ,
\label{eq:6}
\end{equation}
where $\sigma_{0} \equiv \{ \sigma_{0}^1 \ldots \sigma_{0}^L \}$ and
correspondingly for $\sigma_{R}$; the bonds that make the transfer matrices 
${\cal T}_{i}$ belong to $\mathcal{C}$.
For pure systems the $2^L-$component vectors $\tilde\psi$, $\psi$ are determined by the 
boundary conditions; for example, the choice of dominant left and right eigenvectors 
gives the correlation function in an infinite system\cite{fs2}. Here, one needs only be 
concerned with avoiding start-up effects, since there is no convergence of
iterated vectors. This is done by discarding the first few hundred
iterates of the initial vector ${\bf v}_ 0$.
From then on, one can shift the dummy origin of Eq.~(\ref{eq:6}) 
along the strip, taking $\tilde\psi$ ($\psi$) to be the
left-- (right--) iterate of ${\bf v}^{T}_ 0$ (${\bf v}_ 0$~) up to the shifted
origin.  In order to avoid spurious correlations between the dynamical
variables, the respective iterations of $\tilde\psi$ and
$\psi$ must use {\it distinct} realizations of the bond distribution.

As regards conformal invariance, this has been a fundamental tool to the understanding
of the critical behavior of pure systems in two dimensions. Specifically, the
spin-spin correlation function on a strip can be obtained from the conformal mapping
$w=x+iy=(L/2\pi)\ln z$, of the plane $z=u+iv$ onto a strip of width $L$
with periodic boundary conditions across, leading to the following asymptotic behavior
at criticality~\cite{cardy}:
\begin{equation} 
G_{xy}^{\rm pure}\sim \left[ \frac{\pi/L}{\left( \sinh^2 (\pi
x/L)+ \sin^2 (\pi y/L)\right)^{1/2}} \right]^{\eta} \ \ ,
\label{eq:7}
\end{equation}
where $\eta=1/4$ for pure Ising spins.

For disordered spin systems at criticality, one expects
conformal invariance to be preserved when averages over disorder are taken.
Indeed, numerical evidence for this has been found in several instances,
e.g., random-bond Ising model~\cite{dQ95,dQrbs96,dQ97}, random-bond
$q$-state Potts models~\cite{bb&cc}, random transverse Ising
chains at $T=0$ (equivalent to the two-dimensional McCoy-Wu
model)~\cite{prl78_2473}, and Ising spin 
glasses~\cite{hpp01,mc02a,mc02b,prb68_144414,dQ06}.

We first recall the simpler case where correlation decay in the bulk disordered
system is purely polynomial.
Considering correlation functions on a strip, preservation of conformal 
invariance in the sense just defined means that the decay-of-correlation
exponents $\eta_i$, associated to the assorted moments $m_{i}$ of the probability
distribution (PDF) of correlation functions in the disordered magnet, 
can be determined via an extension of Eq.~(\ref{eq:7}), i.e. by assuming:
\begin{equation}
m_{i}\sim z^{-\eta_{i}}\ ,\ z \equiv\left( \sinh^2 (\pi x/L)+ \sin^2 (\pi 
y/L)\right)^{1/2}\ .
\label{eq:8}
\end{equation}
The $\eta_{i}$ can be extracted by fitting numerical data to  
Eq.~(\ref{eq:8}). Thus far, it has been found that results derived in this way 
(e.g. for spin glasses) are consistent with those coming 
from different analytical methods which do not explicitly invoke 
conformal invariance~\cite{hpp01,prb68_144414,dQ06}. 

Before going further, one must recall that an extensive analysis of the
inaccuracies associated to the finite strip width $L$, carried out in 
Ref.~\onlinecite{prb68_144414}, indicates that finite-width effects are
essentially subsumed in the explicit $L$ dependence given in
Eq.~(\ref{eq:7}). Thus, higher-order finite-size corrections presumably do
not play a significant role in the conformal-invariance properties of
assorted moments of the correlation-function PDF (at least when critical correlation
decay in the bulk is purely power law--like).

Turning now to the present case, one must envisage how the multiplicative
logarithmic corrections, predicted in Eqs.~(\ref{eq:1}) and~(\ref{eq:2})
for bulk correlations, translate onto a strip geometry. According to
Refs.~\onlinecite{jlc86,lc87}, for distances $R \gg L$ along the strip
the correlation decay should be exponential, with $m_n \equiv G_n(R) \sim \exp 
(-R/\xi_n)$,
where
\begin{equation}
\frac{L}{\pi\xi_n}=\eta_n -\lambda_n\,\pi\,b\,g(\ln L) + {\cal O} (g^2)\ .
\label{eq:lc1}
\end{equation}
In the above, $b$ is a universal amplitude related to the normalization of the 
three-point function~\cite{jlc86,lc87}, and
\begin{equation}
g(\ln L)=\frac{g_0}{1+\pi\,b\,g_0\,\ln L}\ ,
\label{eq:g}
\end{equation}
where $g_0$ is proportional to the intensity of disorder. In the limit $L \to \infty$,
Eq.~(\ref{eq:lc1}) turns into a $g_0$--independent form,
\begin{equation}
\frac{L}{\pi\xi_n}=\eta_n -\frac{\lambda_n}{\ln L} + {\cal O}
(\frac{1}{\ln L})^2\ ,
\label{eq:lc2}
\end{equation} 
A semi-quantitative verification of Eq.~(\ref{eq:lc2}) was given, for $n=0$ only,
in Ref.~\onlinecite{dQ97}.
 
In Sec.~\ref{results}, we at first apply Eq.~(\ref{eq:8}) to our data, i.e.
logarithmic corrections are not explicitly considered. The idea is to
keep as close as possible to the framework in which conformal invariance
is known to apply to disordered systems, and check whether the 
resulting  effective exponents, $\eta^{\rm eff}_{i}$, deviate from their
pure-system values in a way consistent with the corrections predicted in  
Eqs.~(\ref{eq:1}) and.~(\ref{eq:2}). Then we attempt to extract quantitative
information, by comparing our numerical results to Eq.~(\ref{eq:lc1}), 
as well as
its asymptotic form, Eq.~(\ref{eq:lc2}). Finally, we investigate a phenomenological
scheme to incorporate subdominant corrections to exponential correlation decay on 
a strip, which is subsequently applied to the analysis of low-disorder data.

\section{Numerical Results}
\label{results}

Estimation of uncertainties associated to the several moments of the 
correlation-function PDFs follows the procedure described in
Ref.~\onlinecite{dQrbs96}, where it was shown that for finite-width strips 
of length $N$, the  width of the PDF remains constant
as $N$ varies. Though the dispersions $\Delta G$ (or $\Delta (\ln G)$) 
do not shrink down to zero, it is possible to extract reliable information about 
average values, as the dispersion of these among independent samples is 
proportional to $1/\sqrt{N}$. One can check, for instance, that fluctuations of said 
averages are $\lesssim 1 \%$ for a strip of length $N=10^{6}$.

\begin{figure}
{\centering \resizebox*{3.5in}{!}{\includegraphics*{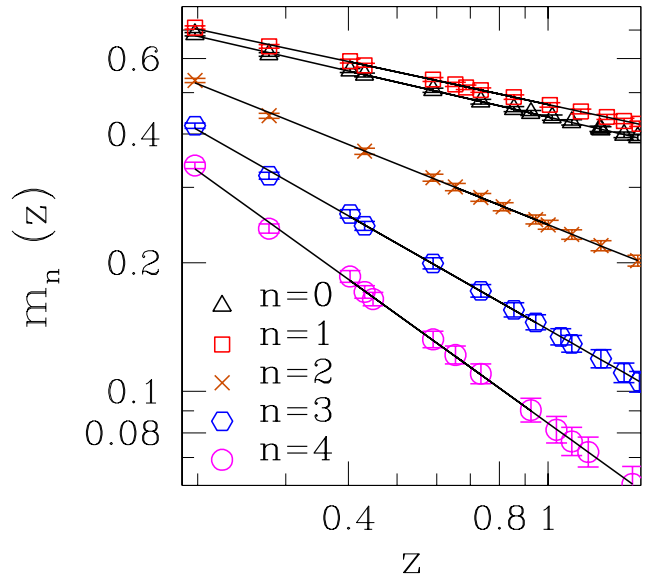}}}
\caption{(Color online) Double-logarithmic plot of moments of the correlation-function 
distribution, 
against $z=\left( \sinh^2 (\pi x/L)+ \sin^2 (\pi y/L)\right)^{1/2}$, for $r=1/4$, 
$L=16$, $N=10^{5}$. 
Lines are least-squares fits of single power-law form to data, from whose slopes one 
extracts estimates for the $\eta^{\rm eff}_{n}$ of
Eq.~(\protect{\ref{eq:8}}).} 
\label{fig:01}
\end{figure}

In Fig. \ref{fig:01}, we present the moments $m_{n}$ of the correlation-function PDF
against the variable $z$, where we used $L=16$ and $N=10^{5}$ for 
$1\leq x \leq 5$, $0 \leq y \leq 8$.  From least-squares fits
of the data, one can find estimates for the effective exponents $\eta^{\rm eff}_{i}$ 
in the form suggested by Eq.~(\ref{eq:8}). Results are displayed in Table~\ref{t1}.
Deviations from pure-system values are consistent with the signs predicted for
logarithmic corrections from Eqs.~(\ref{eq:1}) and ~(\ref{eq:2}), namely an enhancement
for $n=0$, approximate  neutrality for $n=1$, and reduction for $n \geq 2$.

\begin{table}
\caption{\label{t1}
Estimates of effective exponents $\eta^{\rm eff}_{i}$, from least-squares fits of
moments of correlation-function distributions to the form $m_{i}\sim
z^{-\eta^{\rm eff}_{i}}$. 
Data for $r=1/4$, $L=16$ and $N=10^{5}$. 
}
\vskip 0.2cm
\begin{ruledtabular}
\begin{tabular}{@{}llcl}
&$i$& $\eta^{\rm eff}_{i}$ &\\
\hline
& 0 &  0.271(2) &\\
& 1 &  0.252(2) &\\
& 2 &  0.473(4) &\\
& 3 &  0.671(8) &\\
& 4 &  0.85(1) &\\
\end{tabular}
\end{ruledtabular}
\end{table}

In order to 
produce a more quantitative analysis, we now resort to 
Eqs.~(\ref{eq:lc1}),~(\ref{eq:lc2}).

In all cases to be described below, we used  $L=10$, and 
the horizontal distance $x$ was taken long enough in order to
try and detect the effects which give rise to logarithmic 
corrections~\cite{jlc86,lc87}, see Eqs.~(\ref{eq:lc1}),~(\ref{eq:lc2}). In practice, we 
used
$12\lesssim x \lesssim 35$ lattice spacings along the strip, as it was
noticed that the use of larger $x$ gave rise to large sample-to-sample fluctuations,
thus  compromising the accuracy of our data.
Also, and for the sake of simplicity, we made $y=0$. 

Here we evaluated 10 independent estimates for $N=10^{5}$
for the moments of the correlation-function PDF, and took error bars as
three times the standard deviation among these. This was necessary, since it is these 
uncertainties which constitute the single major impediment to the accurate estimation of 
the exponents $\eta_n$ and $\lambda_n$. 

\begin{figure}
{\centering \resizebox*{3.5in}{!}{\includegraphics*{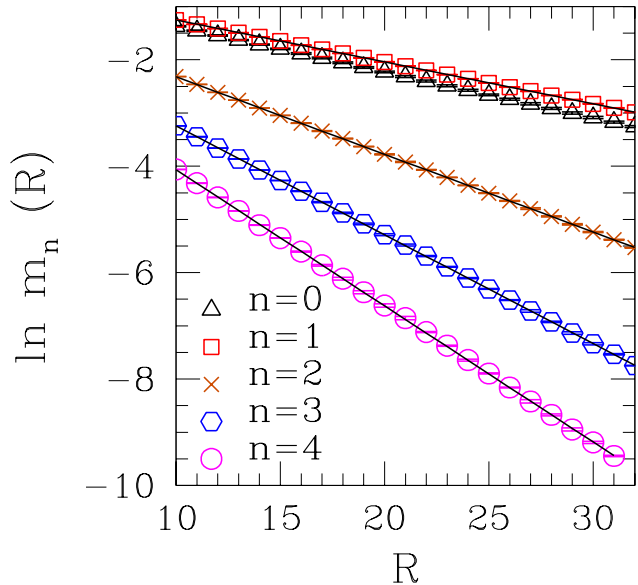}}}
\caption{(Color online) Semi-logarithmic plot of moments of the correlation-function 
distribution, against distance $R$ along the strip (i.e. making $y=0$ in 
Eq.~(\protect{\ref{eq:7}})). Straight lines are least-squares fits of data, from whose 
slopes the assorted values of $\lambda_n$ in Table~\ref{t2} have been estimated
(see text). Here $r=1/4$, $L=10$, $N=10^{5}$.}
\label{fig:cl}
\end{figure}

In the semi-logarithmic plot of Fig.~\ref{fig:cl}, one can see that an exponential
decay with distance, which is a precondition for Eqs.~(\ref{eq:lc1}),~(\ref{eq:lc2}) 
to apply, has set in  to a good degree of accuracy for the range of $R$ depicted. 
We first checked whether the values of strip width $L=10$, and disorder $r=1/4$
used so far, might correspond to the asymptotic regime described by Eq.~(\ref{eq:lc2}).
The corresponding estimates
for $\lambda_n$, shown in Table~\ref{t2}, have been extracted from least-squares fits 
of our data to the form given in Eq.~(\ref{eq:lc2}), by keeping the $\eta_n$ fixed at 
their pure-system values: $\eta_0=1/4$, $\eta_n=n/4\ (n\geq 1)$.

\begin{table}
\caption{\label{t2}
Estimates of exponents $\lambda_{n}$, from least-squares fits of 
moments of correlation-function distributions to a pure-exponential decay form,
as given by Eq.~(\ref{eq:lc2}), keeping the $\eta_n$ fixed at their pure-system 
values: $\eta_0=1/4$, $\eta_n=n/4\ (n\geq 1)$. 
Data for $r=1/4$, $L=10$ and $N=10^{5}$. 
Uncertainties in last quoted digits are shown in parentheses.
}
\vskip 0.2cm
\begin{ruledtabular}
\begin{tabular}{@{}llccl}
&$n$& $\lambda_{n}$\ (calc.) & $\lambda_{n}$\ (expected)&\\
\hline
& $0$ &  $-0.05(1)$ & $-1/8$ &\\
& $1$ &  $-0.0012(7)$ & $0$ &\\
& $2$ &  $0.081(1)$ & $1/4$ &\\
& $3$ &  $0.218(2)$ & $3/4$ &\\
& $4$ &  $0.405(4)$ & $3/2$ &\\
\end{tabular}
\end{ruledtabular}
\end{table}

As can be seen, though the estimates extracted from our data have their sign as
predicted, their magnitude is at most only about one-third of the expected (except 
for  $n=1$, in which case the calculated result may be taken as broadly consistent with 
the predicted absence of corrections). 
The most likely explanation for such discrepancy is, therefore, that the asymptotic
regime has not been reached yet. Similar considerations were used in 
Ref.~\onlinecite{jlc86}, where the $4$-state Potts model was analysed with
help of the generic (i. e., non-asymptotic) form, Eq.~({\ref{eq:lc1}). From 
free-energy data, the numerical values of $g(\ln L)$ for $L=6-10$ 
were extracted, and it was shown that they explained the respective deviations
of gap scaling data from their known exact values. 

A major obstacle to following the same steps here
is the presence of randomness, which severely limits the accuracy of both 
free-energy (see, e g., Ref.~\onlinecite{dQ95}) and correlation-function estimates. 
Furthermore, the exact bulk free energy is not known, contrary to the case of the 
$4$-state Potts model~\cite{jlc86}. 

We then compared  correlation-function data pertaining to different values of $L$, 
namely $L=10$ and $16$ (both for $r=1/4$). We found that, e. g., fitting $L=16$ results 
to the asymptotic form Eq.~({\ref{eq:lc2}) gives estimates for $\lambda_n$ which fall 
essentially within the error bars of those given in Table~\ref{t2} for $L=10$. In 
addition to the randomness effects just mentioned, such
poor resolution also reflects the logarithmic $L$-dependence of the corrections under 
study, and indicates that for the values of $L 
\lesssim 25$ within current reach of transfer-matrix calculations, there is little
hope of extracting reliable information from varying $L$. 
     
Next, we investigated the behavior of our results against varying disorder.
This means varying $r$ in Eq.~(\ref{eq4}), which corresponds to changing $g_0$
in Eq.~(\ref{eq:g}). The actual functional dependence $g_0=g_0(r)$ is 
not known, apart from the basic fact that $g_0$ must increase as $r$ decreases 
towards zero. A similar degree of freedom is absent in the $4$-state Potts model
investigated in Ref.~\onlinecite{jlc86}. From Eqs.~(\ref{eq:lc1}),~(\ref{eq:g}), 
and~~(\ref{eq:lc2}), by doing this for fixed $L$ one expects to move along the 
crossover between small-- and large--$L$ regimes. 

We calculated moments of order $0-4$ of the correlation-function PDFs, for 
$r=0.1$, $0.05$, and $0.01$. Fits of our data to the asymptotic form,
Eq.~(\ref{eq:lc2}), are shown in Fig.~\ref{fig:asympl}. For each $n \neq 1$,
the sign of calculated exponents remains in agreement with predictions and 
their absolute value increases with increasing disorder ($\lambda_1$ remains
close to zero, reaching $0.07(1)$ for $r=0.01$). However, the trend against increasing
disorder is clearly towards overshooting the theoretical predictions, with no sign of
stabilization.

\begin{figure}
{\centering \resizebox*{3.5in}{!}{\includegraphics*{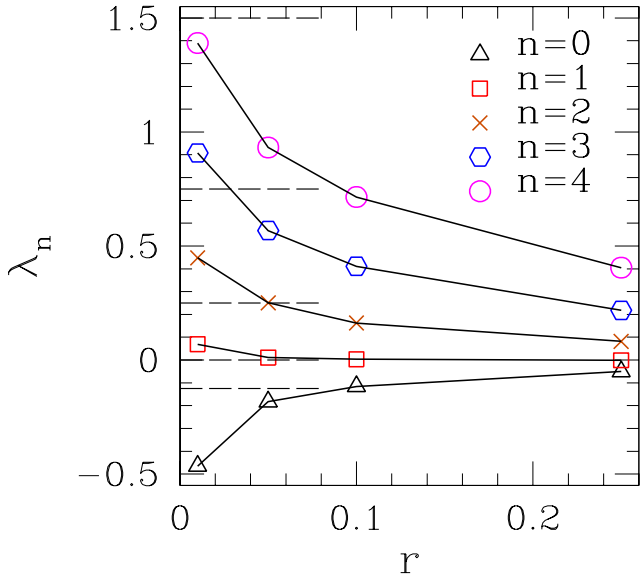}}}
\caption{(Color online) Logarithmic correction exponents $\lambda_n$, 
calculated via the asymptotic expression,
Eq.~(\protect{\ref{eq:lc2}}) from moments of the correlation-function 
distribution, against disorder intensity $r$ (disorder increases with decreasing $r$, 
see Eq.~(\protect{\ref{eq4}})). Error bars are of the same order of, or smaller than,
symbol sizes. Dashed horizontal lines starting on vertical axis
mark expected values of $\lambda_n= -1/8$, $0$, $1/4$, $3/4$, and $3/2$
respectively for $n=0-4$. All for $L=10$, $N=10^{5}$.}
\label{fig:asympl}
\end{figure}

It seems reasonable to exclude the possibility of non-monotonic behavior
for even smaller $r$, which would then bring the estimated exponents back
to their predicted values. The simplest explanation for the behavior depicted
in Fig.~\ref{fig:asympl} is that, at $r \lesssim 0.05$
or thereabouts (for fixed $L=10$), the effects of the percolation fixed point
at $r=0$, $T_c(0)=0$, are beginning to manifest themselves. Indeed,
the description synthesized in Eq.~(\ref{eq:lc1}) implicitly envisages
a perturbation scenario where pure-Ising critical behavior is modified by a
marginally-relevant operator, which induces the logarithmic corrections
investigated here. Thus it is plausible that such framework would fail to 
include the extreme disorder close to the zero-temperature fixed point. 
  
In order to check this hypothesis, we must go back to the non-asymptotic form,
Eq.~(\ref{eq:lc1}), and ask whether (for fixed $L$) there is a range of 
disorder in which the corrections to the pure-system exponents can be decomposed
into the product of an $n$-dependent and an $n$-independent factor  (respectively, 
$\lambda_n$ and $\pi b\,g (\ln L)$) as in Eq.~(\ref{eq:lc1}), and whether such
range is limited by increasing disorder. As pointed out above, in contrast with 
Ref.~\onlinecite{jlc86} here we are allowed to vary $g_0$, which can then substitute 
for the $L$-variation investigated in that Reference.

We assumed the $\eta_n$ and $\lambda_n$ to have their exact predicted values and, from
the numerical estimates of moments of correlation-function PDFs,
calculated $\pi b\,g (\ln L)$ for $n=0$, $2$, $3$, and $4$. The results are exhibited   
in Fig.~\ref{fig:pibg}.
\begin{figure}
{\centering \resizebox*{3.5in}{!}{\includegraphics*{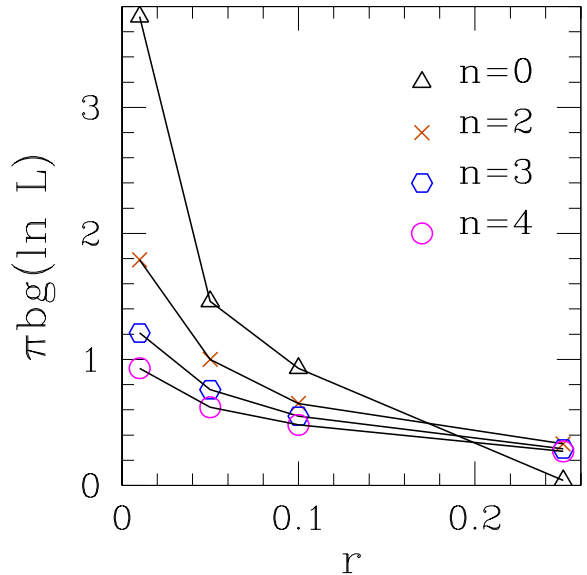}}}
\caption{(Color online) Estimates of $\pi\,b\,g(\ln L)$ of
Eq.~(\protect{\ref{eq:lc2}}) (assuming $\eta_n$, $\lambda_n$ to have their exact values)
calculated from moments of the correlation-function 
distribution, against disorder intensity $r$.
Error bars are of the same order of, or smaller than, symbol sizes.
All for $L=10$, $N=10^{5}$.}
\label{fig:pibg}
\end{figure}
We first note that the $n=0$ data show a much wider range of variation than the rest.
This may be related to specific properties of the zeroth moment of the 
correlation-function PDF, in whose calculation the occurrence of small values plays a 
disproportionately important role (compared to the moments of order $n\geq 1$). 
However, we cannot advance a specific explanation at this point. We decided to 
concentrate on the analysis of $n=2-4$ data, whose average gives $\pi b\,g 
(\ln L)= 0.30(3)$ for $r=1/4$, and $0.56(8)$ for $r=0.1$. For smaller $r$
we see that the scatter becomes larger, in a way consistent with the idea that the
physical picture of Eq.~(\ref{eq:lc1}) becomes frayed as disorder increases.
The agreement is qualitative and, to a fair extent, quantitative, inasmuch as both
Figures~\ref{fig:asympl} and~\ref{fig:pibg} concur in indicating $r \approx 0.1$ 
as the threshold beyond which the perturbative view definitely fails (for $L \approx 
10$, that is; for larger $L$ such threshold would presumably move towards  higher 
disorder).

We now return to the relatively low-disorder case of $r=1/4$, and
investigate whether a perturbative scheme exists, which may account for the features
exhibited by correlation functions on a strip geometry. Specifically,
Eq.~(\ref{eq:lc1}) predicts
that, in the disordered system, correlations on a strip still decay 
purely exponentially, as in pure-Ising strips, 
only with a small change in the exponent. In the
crossover regime to which $L=10$, $r=1/4$ certainly belong, such incipient modification 
may turn up as an effective subdominant (i.e. less than exponential) correction to our 
numerics.
Thus, one could expect an effective behavior:
\begin{equation}
G_{n}(R) \sim \exp(-\pi R\,\eta_n/L)\left[ R \right]^{\Lambda_{n}}\ ,
\label{eq:effecb1}
\end{equation}
where $\eta_n$ are the pure-system exponents, and subdominant corrrections
are assumed to take the simplest possible form of a power law, 
with the $\Lambda_{n}$ to be determined.  
Following this reasoning, and recalling the behavior of bulk correlations
invoked in Eqs.~(\ref{eq:1}) and~(\ref{eq:2}), one sees that a literal
transposition of the $R$-dependence of the latter onto the $z$-dependence of
strip correlations (in analogy with Eq.~(\ref{eq:8})), would mean:
\begin{equation}
G_{n}(z)\sim z^{-\eta_{n}}\left[ \ln z\right] ^{\lambda_{n}}=\left[ z\left(\ln z \right) 
^{q_n}\right]^{-\eta_{n}},
\label{eq:effecb2}
\end{equation} 
where  $\eta_0=1/4,\ \lambda_0=-1/8$, $\eta_{n}=n/4,\ \lambda_{n}=n(n-1)/8\ (n\geq 
1)$, and  $q_n=-\lambda_{n}/\eta_{n}$. In the
regime $x=R \gg L$, $y=0$ considered in Eq.~(\ref{eq:lc1}), which corresponds to
$z \sim \exp (\pi R/L)$ in  Eq.~(\ref{eq:8}), this indeed amounts to writing
\begin{equation}
G_{n}(R) \sim \exp(-\pi R\,\eta_n/L)\left[ R \right]^{\lambda_{n}}\ ,
\label{eq:effecb3}
\end{equation}
i.e., pure-Ising exponential decay with power-law corrections. Since
Eq.~(\ref{eq:effecb2}) does not, as far as we are aware, have a rigorous justification 
on
the basis of conformal invariance, the predicted identification $\Lambda_n=\lambda_n$
extracted from comparison of  Eq.~(\ref{eq:effecb1}) to Eq.~(\ref{eq:effecb3}) must be 
regarded  warily. 
Nevertheless, from the preceding analysis summarized in Tables~\ref{t1} and~\ref{t2},      
for each $n$ one expects $\Lambda_n$ and $\lambda_n$ to have the same 
sign and, possibly, to be of similar orders of magnitude. 
Therefore, in what follows we shall take considerations based on 
Eq.~(\ref{eq:effecb2}) as a starting point
to reanalyze the set of data displayed in Fig.~\ref{fig:cl}.

Using Eq.~(\ref{eq:effecb2}), the moment of order $n=0$  must correspond to  $q=1/2$.
In Fig.~\ref{fig:eta0}, fitting data for the moment of order $n=0$ we at first
fixed $q=1/2$ (as suggested by Eq.~(\ref{eq:2})), and allowed $\eta_0$ to vary in 
Eq.~(\ref{eq:effecb2}). The best fit was found for $\eta_{0}= 0.2505(7)$,
corresponding to a $\chi^2$ per degree of freedom ($\chi^2_{\rm d.o.f.}$) equal to
$0.012$. We then fixed  $\eta_{0}$ at the central estimate just found, and allowed
$\Lambda_0$ of Eq.~(\ref{eq:effecb1}) to vary. The $\chi^2_{\rm d.o.f.}$ for this latter
fit are displayed in the insert of Fig.~\ref{fig:eta0}, with a clear minimum at
$\Lambda_0= -0.125$.

\begin{figure}
{\centering \resizebox*{3.5in}{!}{\includegraphics*{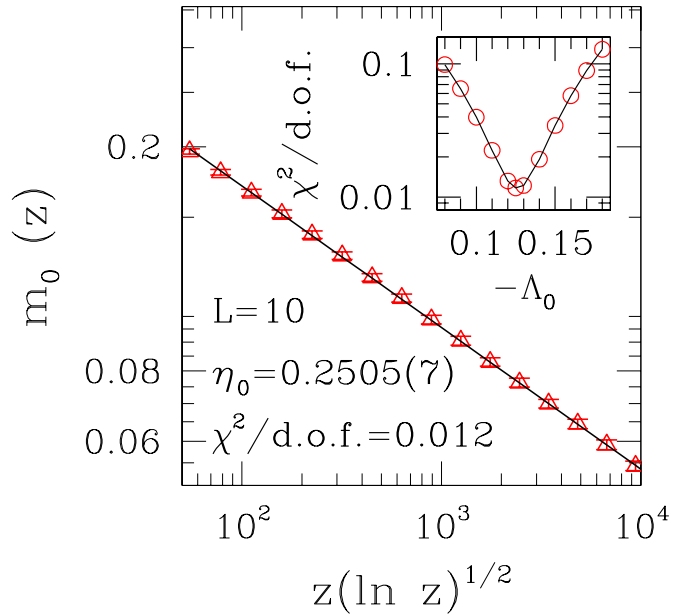}}}
\caption{(Color online) Main diagram: double-logarithmic plot of moment of
order $n=0$ of correlation-function PDF, against
$\left[ z\left( \ln z \right) ^{q}\right]$ for $q=1/2$. Line corresponds to
 $\eta_0= 0.2505$ (best estimate obtained for fixed $q=1/2$, see text); $L=10$, $r=1/4$.
Insert: semi-logarithmic plot of $\chi^2_{\rm d.o.f.}$ against  $-\Lambda_0$, 
for fits of $n=0$
moment data against $z^{-\eta_{0}}\left[ \ln z\right] ^{\Lambda_{0}}$, with $\eta_0= 
0.2505$ (fixed). } 
\label{fig:eta0}
\end{figure}
According to Eq.~(\ref{eq:1}), the moment of order $n=1$ is  not expected to present 
logarithmic corrections. In order to check this,  we fixed $q=0$ in
Eq.~(\ref{eq:effecb2}) and allowed $\eta_1$ to vary.
The best fit, shown in Fig.~\ref{fig:eta1}, was found for $\eta_{1}= 0.2505(1)$, very 
close to $1/4$ of a uniform  system, corresponding  to a $\chi^2_{\rm d.o.f.}=0.02$. 
In the insert of the Figure, one can see that, upon keeping $\eta_1$ fixed at the 
central estimate quoted above, and allowing
$\Lambda_1$ of Eq.~(\ref{eq:effecb1}) to vary, the closest fit indeed occurs for 
$\Lambda_1=0$.
\begin{figure}
{\centering \resizebox*{3.5in}{!}{\includegraphics*{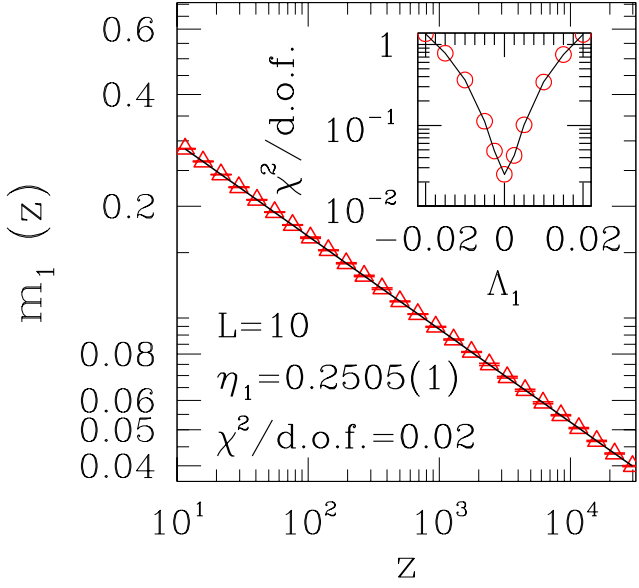}}}
\caption{(Color online) Main diagram: double-logarithmic plot of moment of
order $n=1$ of correlation-function PDF, against
$\left[ z\left( \ln z \right) ^{q}\right]$ for $q=0$. Line corresponds to
 $\eta_1= 0.2505$ (best estimate obtained for fixed $q=0$, see text); $L=10$, $r=1/4$.
Insert: semi-logarithmic plot of $\chi^2_{\rm d.o.f.}$ against  $\Lambda_1$ , for fits 
of $n=1$ moment data against $z^{-\eta_{1}}\left[ \ln z\right]
^{\Lambda_{1}}$, with $\eta_1= 0.2505$ (fixed). } 
\label{fig:eta1}
\end{figure}

When turning to higher-order moments $n \geq 2$, we noticed that it was not possible,
in general, to produce good fits of the whole range of data displayed in 
Fig.~\ref{fig:cl} to the variables suggested by Eq.~(\ref{eq:effecb2}). However, subsets
of the data did fit well against such variables, as is illustrated in 
Fig.~\ref{fig:eta234}. We shall return to this point at the end of the Section.

\begin{figure}
{\centering \resizebox*{3.5in}{!}{\includegraphics*{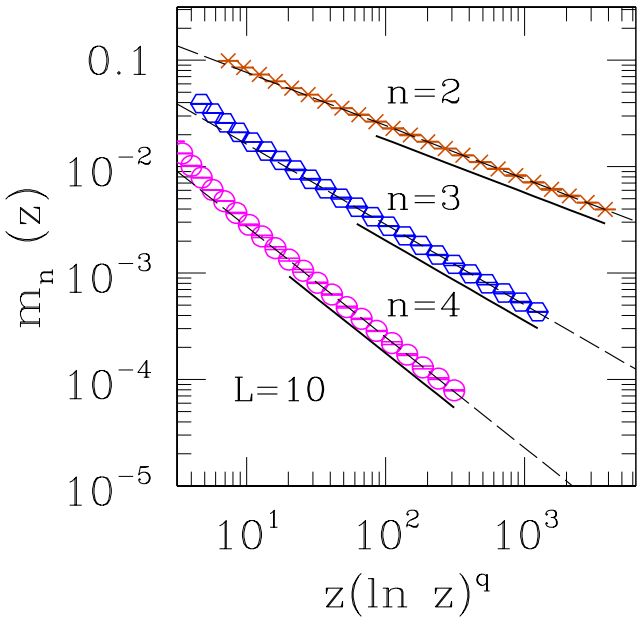}}}
\caption{(Color online) Double-logarithmic plot of higher-order moments of
correlation-function PDF, against
$\left[ z\left( \ln z \right) ^{q}\right]$ with $q=-1/2,$ $-1$, and $-3/2$
respectively for $n=2$, $3$, and $4$.
Thick continuous lines show range of data used in least-squares fits of
the $\eta_n$ for $q$ fixed as above. $L=10$, $r=1/4$.} 
\label{fig:eta234}
\end{figure}

For the moment of order $n=2$, again we 
fixed $q=-1/2$ in Eq.~(\ref{eq:effecb2}) and allowed $\eta_2$ to vary. 
The best fit was found for $\eta_{2}= 0.4992(8)$,
corresponding to a $\chi^2_{\rm d.o.f.}=0.03$. Data for this setup are displayed in
the main diagram of Fig.~\ref{fig:eta2}. In the insert of the Figure, we kept $\eta_{2}$ 
fixed at the central estimate just found, and allowed
$\Lambda_2$ of Eq.~(\ref{eq:effecb1}) to vary. The $\chi^2_{\rm d.o.f.}$ of the 
respective
fits are shown, with a clear minimum at $\lambda_2= 0.25$.
\begin{figure}
{\centering \resizebox*{3.5in}{!}{\includegraphics*{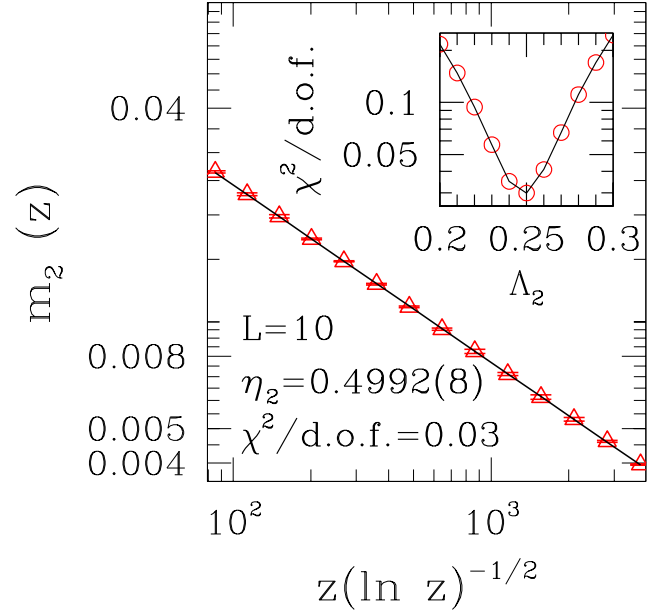}}}
\caption{(Color online) Main diagram: double-logarithmic plot of moment of
order $n=2$ of correlation-function PDF, against
$\left[ z\left( \ln z \right) ^{q}\right]$ for $q=-1/2$. Line corresponds to
 $\eta_2= 0.4992$ (best estimate obtained for fixed $q=-1/2$, see text); $L=10$, 
$r=1/4$.
Insert: semi-logarithmic plot of $\chi^2_{\rm d.o.f.}$ against  $\Lambda_2$, 
for fits of $n=2$
moment data against $z^{-\eta_{2}}\left[ \ln z\right] ^{\Lambda_{2}}$, with $\eta_2= 
0.4992$ (fixed). } 
\label{fig:eta2}
\end{figure}

The previous procedure was applied for the moments of orders $n=3$ and $n=4$. 
The results for all investigated exponents, $\eta_{n}$ and $\Lambda_{n}$ are 
displayed in Table~\ref{t3}.
\begin{table}
\caption{\label{t3}
Estimates of exponents $\eta_{n}$ and $\Lambda_{n}$, from least-squares fits of
moments of correlation-function distributions obtained with 
$G_{n}(z)\sim \left[ z\left(\ln z \right) 
^{q_n}\right] ^{-\eta_{n}}$, as given by Eq.~(\ref{eq:effecb2}); see also 
Eq.~(\ref{eq:effecb1}). Data for $r=1/4$, $L=10$ and $N=10^{5}$. 
Uncertainties in last quoted digits are shown in parentheses.
}
\vskip 0.2cm
\begin{ruledtabular}
\begin{tabular}{@{}llccl}
&$n$& $\eta_{n}$ & $\Lambda_{n}$&\\
\hline
& 0 &  $0.2505(7)$ & $-0.124(4)$ &\\
& 1 &  $0.2505(1)$ & $0.000(2)$ &\\
& 2 &  $0.4992(8)$ & $0.248(6)$ &\\
& 3 &  $0.751(2)$ & $0.75(1)$ &\\
& 4 &  $1.032(6)$ & $1.54(4)$ &\\
\end{tabular}
\end{ruledtabular}
\end{table}

It can be seen from Table~\ref{t3} that both accuracy and
precision  (the latter, upon comparison with theoretical predictions of $\lambda_n$) of 
results become lower for  higher-order moments; in particular, our estimate 
of $\eta_4$ already differs somewhat from $\eta_4 =1$ predicted in
Eq.~(\ref{eq:2}) (though that for $\Lambda_4$ is consistent, at the margin,
with $\lambda_4=3/2$ from the same Equation). 
For $n>4$ we have found that the situation worsens 
considerably. This is because: (i) short-distance deviations from 
asymptotic behavior tend to increase with $n$, and (ii) 
at the relatively long distances where asymptotic behavior sets in, the
correlation functions themselves, and {\em a fortiori} 
their higher moments, become smaller, and their averages prone to higher
relative  fluctuations. We recall the trend illustrated in Figure~\ref{fig:eta234}, 
against increasing $n$: for all three sets of data shown there, the
thick continuous lines (corresponding respectively to $19 \leq x \leq 32\ (n=2)$, $21 
\leq x \leq 32\ (n=3)$, and  $20 \leq x \leq 31\ (n=4))$ denote the range of data used
to extract the results quoted in Table~\ref{t3}. One sees that, since $|q|$
increases with $n$, sets of data for the same range of $x$ become more
compressed when plotted against $\left[ z\left( \ln z
\right) ^{q}\right]$, thus increasing uncertainties upon fitting.  

As one might expect, such a picture of good fits to Eq:~(\ref{eq:effecb1}), producing 
estimates of $\eta_n$ and 
$\lambda_n$ quite close to the exact values, does not hold for higher disorder. For 
instance, already for $r=0.1$, $n=2$, our best fit to Eq.~(\ref{eq:effecb1}) gives
 $\eta_2=0.433(1)$, $\lambda_2=0.019(5)$ ($\chi^2_{\rm d.o.f.}=0.04$). Fixing 
$\eta_2=1/2$, $\lambda_2=1/4$ gives $\chi^2_{\rm d.o.f.} \approx 24$.
  

\section{Conclusions} 
\label{sec:conc}
We have tested the theoretical predictions, given in Eqs.~(\ref{eq:1}) and~(\ref{eq:2}), 
of logarithmic corrections to the asymptotic behavior of spin-spin correlation 
functions, in the two-dimensional (unfrustrated) random-bond Ising model.
The use of a strip geometry has been accompanied by the corresponding
finite-size scaling and conformal-invariance considerations. So far the
validity of the latter, when dealing with non-uniform systems, refers to 
 disorder-averaged quantities, and has been  established only in cases where 
critical correlation decay 
is purely polynomial~\cite{hpp01,prb68_144414,dQ06}.

From examination of short-distance data (see Fig.~\ref{fig:01}, and  Table~\ref{t1}), 
we could only establish that the  effective exponents obtained were  consistent with 
corrections of the same sign as 
predicted by theory. The situation is similar to that found in Ref.~\onlinecite{bs04}, 
where the conformal mapping of the square lattice on a finite square was used, 
and only $n=0$ was investigated . Those authors concluded that, for the size of lattices 
used, they would be able only to assert that the sign of corrections was as predicted.

By considering suitably long distances along the strip, we attempted to reach
the regime where the effect of logarithmic corrections on a strip geometry
is expected to be more easily isolated, as predicted in Refs.~\onlinecite{jlc86,lc87}.
Allowing the intensity of disorder to vary produced a picture in which
one is able to identify the approximate range of validity of the results given 
in Refs.~\onlinecite{jlc86,lc87}, even for strips of relatively small width.
This relies essentially on identifying the interval of disorder strength $r$,
along which an approximately $n$-independent factor can be isolated from numerical 
data, in accordance with the non-asymptotic forms predicted in 
Eq.~(\ref{eq:lc1}). The corresponding data are shown in Fig.~\ref{fig:pibg}.

We also produced a phenomenological fitting scheme, expected to work at low disorder,
and synthesized in Eq.~(\ref{eq:effecb1}),
from which estimates of the postulated exponents $\Lambda_n$ were extracted.
While the analogy to Eq.~(\ref{eq:8}), used to establish Eqs.~(\ref{eq:effecb2})
and~(\ref{eq:effecb3}), seems short of a rigorous justification on the basis of
conformal invariance, the ensuing prediction that $\Lambda_n=\lambda_n$
appears to be backed to a large extent by our numerical data for low disorder $r=1/4$.
As expected, the validity of such a scheme does not extend far towards increasing 
disorder. 

In summary, the evidence shown here provides a check of the validity of 
Eqs.~(\ref{eq:1}) and~(\ref{eq:2}) for disordered Ising systems, in the 
appropriate limit of long distances. It would be interesting to ascertain
whether any deeper connection lies between the phenomenological form 
Eq.~(\ref{eq:effecb1})
and conformal-invariance properties.
\begin{acknowledgments}
J.C.L. thanks the Brazilian agency CAPES for partial financial support, and
Departamento de F\'\i sica de S\'olidos, UFRJ, for making its research
infrastructure available.    
The research of S.L.A.dQ. was partially supported by
the Brazilian agencies CNPq (Grant No. 30.0003/2003-0), FAPERJ (Grant
No. E26--152.195/2002), FUJB-UFRJ, and Instituto do Mil\^enio de
Nanoci\^encias--CNPq.
\end{acknowledgments}

\end{document}